\documentstyle[amssymb,prl,aps,epsfig]{revtex}

\begin{document}
\title{Phonon-mediated electron spin phase diffusion in a quantum dot}
\author{Y. G. Semenov and K. W. Kim}
\address{Department of Electrical and Computer Engineering\\
North Carolina State University, Raleigh, NC 27695-7911}
\maketitle

\begin{abstract}
An effective spin relaxation mechanism that leads to electron spin
decoherence in a quantum dot is proposed. In contrast to the
common calculations of spin-flip transitions between the Kramers
doublets, we take into account a process of phonon-mediated
fluctuation in the electron spin precession and subsequent spin
phase diffusion. Specifically, we consider modulations in the
longitudinal g-factor and hyperfine interaction induced by the
phonon-assisted transitions between the lowest electronic states.
Prominent differences in the temperature and magnetic field
dependence between the proposed mechanisms and the spin-flip
transitions are expected to facilitate its experimental
verification. Numerical estimation demonstrates highly efficient
spin relaxation in typical semiconductor quantum dots.
\end{abstract}

\address{Department of Electrical Computer Engineering\\
North Carolina State University, Raleigh, NC 27695-7911}

Recently, there has been much interest in the study of electron
spin dechoherence mechanisms in quantum dots (QDs) since they are
a natural candidate for the qubit operations of quantum computing.
A typical approach to this problem is to calculate the spin
transition probability caused by the electron spin interaction
with a thermal bath~\cite{KhaetNaz01,Tahan,GlavinKim}. A
thermostat composed of nuclear spins is not an effective
dissipative system due to the long nuclear spin relaxation:
however, the nuclei can assist in the phonon
relaxation~\cite{ErlNazFal,Merkulov,KhaetLoss,SemKim}.

Since spin-lattice relaxation (SLR) is assumed to be associated
with spin-flip transitions between different electron spin states,
the non-diagonal spin-flip matrix elements must be taken into
account. Most studies of SLR in the literature have analyzed
various mechanisms responsible for such transitions. Specifically,
there are two approaches to this problem. The first is based on
the spin-orbital interaction which is particular to
III-V~\cite{KhaetNaz01} or Si/Ge~\cite{Tahan,GlavinKim} QDs. The
second approach incorporates the nuclear hyperfine interaction
(HFI)~\cite{Merkulov,KhaetLoss,SemKim} as a factor,
leaving the time-reversal symmetry for the electron spin Hamiltonian~\cite%
{ErlNazFal}.

Actually, there are no principal differences between the
aforementioned approaches to the SLR in QDs and the early
pioneering works performed in the 1960's for shallow donor
relaxation (with the exception of the specific electronic energy
structure and the influence of strain). Both the early
works~\cite{Hasegawa,Roth,WilsonFeher} and the  later studies of
QD SLR~\cite{KhaetNaz01,Tahan,GlavinKim} deal with the {\em
longitudinal} (or energy) relaxation accompanying the exchange
between the Zeeman and phonon reservoirs. Note, however, that
quantum computing is qualitatively limited by other relaxation
processes which result in the destruction of the electron spin
phase coherence that can occur without energy relaxation.
Development of {\em transversal} (or phase) relaxation was not the
goal of the early studies nor the recent SLR research in QDs.

In this paper, we show that there are decoherence mechanisms specific to
phase relaxation, which can be rather more effective than energy relaxation.
To make this assertion more clear, let us consider an electron
spin $\vec{s}$ under
the influence of a magnetic field directed along the $z$ axis
with a randomly fluctuating strength.
In this case, the projection of electron spin
on the z axis $s_{z}$ is conserved and no longitudinal relaxation occurs.
Nevertheless, the phase of electron spin will change randomly with the Zeeman
frequency fluctuation $\delta \Omega $ resulting in
a decoherence rate of $T_{2}^{-1}\sim \delta \Omega ^{2}\tau _{c}$.
Here, $\tau _{c}$ is the correlation time of these
fluctuations~\cite{Sem03}. This leads to a very important conclusion that
the phase relaxation time may not be determined by the matrix elements
between the Kramers doublets, the most important
restriction on the longitudinal SLR.

In order to calculate the fluctuations in the Zeeman frequency $\Omega$,
we will consider the
phonon-induced transitions between the lowest electronic states $\left\vert
k\right\rangle $ which possess different spin splitting and are not linked by
the time reversal operator.
The resulting electron spin phase diffusion due to the spin
precession in a fluctuating field is expected to be an efficient relaxation
mechanism if (i) the electron significantly changes spin precession
frequencies when the transitions between different states $\left\vert
k\right\rangle $ occur and (ii) the mean time $\tau _{c}$ of phonon assisted
transitions between these states is less than spin relaxation time $T_{2}$
\cite{Sem03}. In the case of a QD, these two conditions can be easily
realized due to shallow energy levels and the $g$-factor or hyperfine
constant dependence on the orbital electronic states.

We begin the quantitative analysis by defining the Hamiltonian $H$ over
the basic functions reduced to a few lowest electronic states $\left\vert
k\right\rangle $, which are involved due to the phonon-assisted transitions.
We also assume that the single-electron problem in a QD without Zeeman
energy and HFI gives the doubly degenerate energy spectrum $E_{k}$ with
eigenstates $\left\vert k\right\rangle $. It is conveniently (but not
necessary) assumed that the spin splitting $\left\vert \Omega \right\vert $
is small with respect to the energy intervals $\left\vert E_{k}-E_{k^{\prime
}}\right\vert $. With this assumption, the total Hamiltonian takes the form%
\begin{equation}
H=H_{s}+H_{e}+H_{ph}+H_{e-ph}.  \label{eq1}
\end{equation}

The first term $H_{s}=\vec{\Omega}\vec{s}$ is the spin (or pseudospin)
energy Hamiltonian; in the most general case, the frequency operator reads%
\begin{equation}
\vec{\Omega}=%
\mathop{\displaystyle\sum}%
\limits_{k,k^{\prime }}\vec{\Omega}_{k,k^{\prime }}\left\vert k\right\rangle
\left\langle k^{\prime }\right\vert ,  \label{q2}
\end{equation}%
where $\vec{\Omega}_{k,k^{\prime }}$ are the matrix elements of the
effective field (in units of energy) taken between the $\left\vert
k\right\rangle $ and $\left\vert k^{\prime }\right\rangle $ states. The
spin-independent electron energies describe the Hamiltonian $ H_e$ %
\begin{equation}
H_{e}=%
\mathop{\displaystyle\sum}%
\limits_{k}E_{k}\left\vert k\right\rangle \left\langle k\right\vert .
\label{eq3}
\end{equation}%
The Hamiltonians of the lattice and electron-phonon interactions have the
usual form%
\begin{eqnarray}
H_{ph} &=&%
\mathop{\displaystyle\sum}%
\limits_{q}\omega _{q}(a_{q}^{+}a_{q}+%
{\frac12}%
);  \label{eq4} \\
H_{e-ph} &=&%
\mathop{\displaystyle\sum}%
\limits_{q,k,k^{\prime }}B_{k,k^{\prime }}^{q}\left\vert k\right\rangle
\left\langle k^{\prime }\right\vert (a_{q}^{+}+a_{-q}).  \label{eq5}
\end{eqnarray}%
Here $q=\{\vec{q},\varkappa \}$ represents the wave vector and
polarization of a phonon with energy $\omega _{q}$; $-q\equiv
\{-\vec{q},\varkappa \}$, $a_{q}^{+}$ and $a_{q}$ are the phonon
creation and annihilation operators, $B_{k,k^{\prime }}^{q}$ is
the matrix element of the electron-phonon interaction, which
depends on the material parameters and the geometry of the QD. In
Eq.\ (\ref{eq1}), the last three terms constitute the Hamiltonian
of the dissipative sub-system responsible for electron spin
relaxation, $H_{d}=H_{e}+H_{ph}+H_{e-ph}$.

We are interested in the evolution of electron
spin $\vec{s}(t)=Tr\{\rho (t)\vec{s}\}$ ($\rho (t)$ is a density matrix)
in a system with the Hamiltonian of Eq.\ (\ref{eq1}).
This evolution was shown to be described by the quantum kinetic equation
(see Ref.\cite{Sem03})%
\begin{equation}
\frac{d}{dt}\vec{s}(t)=\vec{\omega}\times \vec{s}(t)-{\bf \Gamma }\left(
\vec{s}(t)-\vec{s}_{0}\right) ,  \label{eq6}
\end{equation}%
where $\vec{\omega}=\left\langle \vec{\Omega}\right\rangle $ is an
effective magnetic field, $\left\langle \ldots \right\rangle
=Tr\{e^{-H_{d}/T}\ldots \}/Tre^{-H_{d}/T}$, and $T$ is the
temperature; $\vec{\omega}$ and $T$ are in units of energy. The
matrix ${\bf \Gamma }$ of relaxation coefficients is composed of
Fourier transformed correlation functions $\gamma _{\mu \nu
}\equiv \gamma _{\mu \nu }\left( \omega \right) $=$\left\langle
\delta \Omega _{\mu }\left( \tau \right) \delta \Omega _{\nu
}\right\rangle _{\omega }$=$\frac{1}{2\pi }\int\limits_{-\infty
}^{\infty }\left\langle \delta \Omega _{\mu }\left( \tau \right)
\delta \Omega _{\nu }\right\rangle
e^{i\omega \tau }d\tau $; $\mu $, $\nu =x$, $y$, $z$; $\delta \vec{\Omega}=%
\vec{\Omega}-\left\langle \vec{\Omega}\right\rangle $. With a
provision that the correlation functions are symmetrical, $\gamma
_{\mu \nu }\left( \omega \right) =\gamma _{\nu \mu }\left( \omega
\right) $, the matrix ${\bf \Gamma =\pi }\left\Vert \Gamma _{\mu
\nu }^{\prime }\right\Vert $ has a simple form
in the frame of references related to the direction of the effective field $%
\vec{\omega}\parallel \hat{z}$: $\Gamma _{xx}^{\prime }=\gamma
_{zz}^{0}+n\gamma _{yy}$, $\Gamma _{yy}^{\prime }=\gamma _{zz}^{0}+n\gamma
_{xx}$, $\Gamma _{zz}^{\prime }=n(\gamma _{xx}+\gamma _{yy})$, $\Gamma _{\mu
\nu }^{\prime }=-n\gamma _{\mu \nu }$, ($\mu \neq \nu $), where $\gamma
_{zz}^{0}=\gamma _{zz}(0)$, $n\equiv n(\omega )=\left( 1+e^{\omega
/T}\right) /2$, $\vec{s}_{0}=-%
{\frac12}%
\tanh (\omega /2T)\{0,0,1\}$.

Thus, the problem of spin relaxation is reduced to the calculation of the
correlation functions of the effective field operator with the Hamiltonian $%
H_{d}$ of the dissipative subsystem. These calculations strongly depend on
the specific form of $H_{d}$, the energy spectrum, and the quantity of
electron states considered. Keeping this context in mind, we consider the
more simple problem of electron fluctuations between only two discreet
states $\left\vert k\right\rangle =\left\vert g\right\rangle $ or $%
\left\vert e\right\rangle $ corresponding to the ground state and the first
exited (by interval $\delta $) electronic energy level with Zeeman
frequencies $\vec{\Omega}^{g}$ and $\vec{\Omega}^{e}$. Such a simplification
allows us to easily perform all the necessary calculations in an analytical
form. In addition, most of the important physics of the new mechanism under
consideration can be obtained in the framework of this two-level model.

Hereafter, it is convenient to introduce Pauli matrices $\sigma _{1}$, $%
\sigma _{2}$, $\sigma _{3}$ on the basis $\left\vert e\right\rangle $, $%
\left\vert g\right\rangle $, where according to definition, $\sigma _{1}$, $%
\sigma _{2}$, $\sigma _{3}$ are invariant with respect to the coordinate system
rotation in contrast to actual spin matrices $\vec{s}$. In terms of Pauli
matrices, the Hamiltonian of dissipative subsystem takes the form $%
H_{d}=H_{ph}+\delta \sigma _{3}+\Sigma _{q}B_{q}\sigma _{1}(a_{q}^{+}+a_{-q})
$. The electron spin Hamiltonian now assumes the form $H_{s}=%
{\frac12}%
(\vec{\Omega}^{e}+\vec{\Omega}^{g})\vec{s}+%
{\frac12}%
(\vec{\Omega}^{e}-\vec{\Omega}^{g})\sigma _{3}\vec{s}$, which defines the
fluctuating part of the effective field $\delta \vec{\Omega}=%
{\frac12}%
(\vec{\Omega}^{e}-\vec{\Omega}^{g})(\sigma _{3}-\left\langle \sigma
_{3}\right\rangle )$ with $\left\langle \sigma _{3}\right\rangle =-\tanh
\left( \delta /2T\right) $ and gives the correlation functions in Eq.\ (\ref%
{eq6}) in the form
\begin{eqnarray}
\gamma _{\mu \nu }\left( \omega \right)  &=&%
{\frac14}%
(\Omega _{\mu }^{e}-\Omega _{\mu }^{g})(\Omega _{\nu }^{e}-\Omega _{\nu
}^{g})J_{\omega }(T);  \label{eq8} \\
J_{\omega }(T) &=&\left\langle \left( \sigma _{3}\left( \tau \right)
-\left\langle \sigma _{3}\right\rangle \right) \left( \sigma
_{3}-\left\langle \sigma _{3}\right\rangle \right) \right\rangle _{\omega }.
\label{eq9}
\end{eqnarray}

Skipping the derivation, we present the final result for the case $\omega
\ll \delta $,
\begin{eqnarray}
J_{\omega }(T) &=&\frac{1-\left\langle \sigma _{3}\right\rangle ^{2}}{\pi
n(\omega )}\frac{\tau _{c}}{\omega ^{2}\tau _{c}^{2}+1};  \label{eq10} \\
\tau _{c}^{-1} &=&2\pi
\mathop{\displaystyle\sum}%
\limits_{q}\left\vert B_{q}\right\vert ^{2}(2n_{q}+1)\delta \left( \omega
_{q}-\delta \right) ,  \label{eq11}
\end{eqnarray}%
where $n_{q}=\left\langle a_{q}^{+}a_{q}\right\rangle $ is the phonon
population factor for mode $q$. The parameter $\tau _{c}$ has the simple
physical meaning of the correlation time caused by phonon-assisted
transitions between the $g$- and $e$-states.

Actually, Eqs.\ (\ref{eq8})-(\ref{eq11}) describe the problem under
consideration in a very general form. Before we specify the electron spin interaction,
which fluctuates due to the phonon induced transitions between $\left\vert
g\right\rangle $ and $\left\vert e\right\rangle $ states, we provide the
analysis of the SLR temperature dependence. In doing so, we note that the
correlation time $\tau _{c}$ given in Eq.\ (\ref{eq11}) can be written as $\tau _{\delta
}\tanh \left( \delta /2T\right) $, where $\tau _{\delta }=\left[ 2\pi \Sigma
_{q}\left\vert B_{q}\right\vert ^{2}\delta \left( \omega _{q}-\delta \right) %
\right] ^{-1}$ is the lifetime of the excited electron state with respect to
the transition to the ground state through phonon emission in the limit $%
T\rightarrow 0$. Thus, at zero magnetic field (i.e., $\omega =0$) the
temperature dependence of SLR is reduced to
\begin{equation}
J_{0}(T)=\frac{\tau _{\delta }}{\pi }F\left( \frac{\delta }{2T}\right) ; ~~~
F\left( x\right) =\left( 1-\tanh ^{2}x\right) \tanh x.
\label{eq12}
\end{equation}%
When the magnetic field is strong enough, this factor also describes the
temperature dependence of the transversal relaxation because only $\gamma
_{zz}^{0}$ survives in the matrix ${\bf \Gamma }$ [Eq.\ (\ref{eq6})]
in the limit $ \omega ^{2}\tau _{c}^{2}\gg 1$.

As shown in Fig.\ 1, the pronounced maximum in the temperature
dependence of $F\left( \delta/2T\right) $
around $T=\delta $ has a simple physical meaning. The left of the peak
corresponds to the reduced hopping from the $\left\vert
g\right\rangle $ to $\left\vert e\right\rangle $ state that decreases the
difference $\vec{\Omega}^{g}-\left\langle \vec{\Omega}^{g}\right\rangle $
(or the amplitude of fluctuations). So, in the limit $T\ll \delta $ the
fluctuations are frozen out and our mechanism becomes non-effective as $\exp
(-\delta /T)$. The negative slope in the right side (high temperature)
arises due to the
well-known effect of fluctuation dynamical averaging, which becomes more
pronounced with an increase in temperature .

As noted above, the correlation function $\pi \gamma _{zz}^{0}=\frac{\pi }{4}%
(\Omega _{z}^{e}-\Omega _{z}^{g})^{2}J_{0}(T)$ describes the rate $T_{2}^{-1}
$ of the spin relaxation if longitudinal fluctuations dominate over
transversal ones, $\gamma _{zz}^{0}\gg \gamma _{xx}(\omega )$, $\gamma
_{yy}(\omega )$. Let us apply the general theory discussed above to the
mechanism of phase relaxation, which stems from the hopping between excited
and ground states with different $g$ factors. In the most general case,
the reason for such a difference is the $g$-factor dependence on the energy
separation between the electron's discrete level and nearest spin-orbital
split electronic band. For technologically significant III-V compounds,
where the interaction with the valence band edge determines the deviation of
electron $g$-factor from free electron Land\`{e} factor $g_{0}\approx 2$
\cite{KiselevIvch}, one can find the amplitude of the fluctuation $\Delta
g=\delta (g_{0}-g)(\Delta _{so}+2E_{g})/E_{g}(\Delta _{so}+E_{g})$, where $%
E_{g}$ is the band gap, and $\Delta _{so}$ the spin-orbital
splitting of the valence band; we also assume inequality $\delta
\ll $ $E_{g}$. In the case of a silicon QD,  $E_{g}$ is a
splitting $E_{15}$ of $\Delta $ point in the Brillouin
zone~\cite{Roth}. The relaxation mechanism due to the $g$-factor
anisotropy of the $\Delta $ band is not effective in the case of
Si QD because of the specific valley orbital
structure~\cite{GlavinKim}. The final equation for the
phonon-assisted rate of phase relaxation caused by the Zeeman
energy fluctuations is given by%
\begin{equation}
T_{2,Z}^{-1}=\frac{(g_{0}-g)^{2}}{4g^{2}}\left( \frac{\delta (\Delta
_{so}+2E_{g})}{E_{g}(\Delta _{so}+E_{g})}\right) ^{2}\omega _{0}^{2}\tau
_{\delta }F\left( \frac{\delta }{2T}\right) ,  \label{eq13}
\end{equation}%
where $\hbar \omega _{0}=g\mu _{B}B$, and $\mu _{B}$ is the Bohr magneton.
One can see that our mechanism reveals a quadratic dependence of $%
T_{2,Z}^{-1}$ on the applied magnetic field $B$ in contrast to the $B^{4}$%
-dependence found in the previous calculations of longitudinal
SLR~\cite{KhaetNaz01,Tahan,GlavinKim,Hasegawa,Roth}

An estimation of excited state lifetime $\tau _{\delta }$ can be performed
in terms of a deformation potential interaction and a model of lateral
carrier confinement~\cite{GlavinKim}. The matrix element of electron-phonon
interaction between the $\left\vert g\right\rangle $ and $\left\vert
e\right\rangle $ states in this model is $B_{q}=iC\sqrt{\hbar q/2\rho
v_{\parallel }V_{0}}J_{osc}$, where $J_{osc}=J_{osc}(\vec{q})$ is a
corresponding form-factor calculated in Ref.\cite{GlavinKim}, $C$ is the
deformation potential, and $\rho $, $v_{\parallel } $ and $V_{0}$ are the
density, longitudinal sound velocity, and volume of crystal, respectively. A
straightforward calculation of inverse lifetime results in the expression%
\begin{equation}
\tau _{\delta }^{-1}=\frac{C^{2}q_{\delta }^{3}\alpha }{32\pi ^{2}\hbar \rho
v_{\parallel }^{2}}%
\displaystyle\int %
\limits_{0}^{1}(1-z^{2})e^{-\alpha (1-z^{2})}dz,  \label{tau}
\end{equation}%
where $q_{\delta }=\delta /\hbar v_{\parallel }$, $\alpha =\delta
/2m_{e}v_{\parallel }^{2}$, and $m_{e}$ is a lateral effective mass.

To show the efficiency of the mechanism under consideration, we assume $%
\delta =2 $ meV\cite{GlavinKim} and calculate the relaxation parameters
of Eqs.\ (\ref%
{tau}) and (\ref{eq13}) for a GaAs QD under the magnetic field $B=1$ T. We
find that $\tau _{c}\simeq \tau _{\delta }=5.8 \times 10^{-9}$ s, and $%
T_{2}=2.0$ s, $1.8\times 10^{-5}$ s and $5.4\times 10^{-8}$ s for $T=1$ K, $2$
K, and $4$ K, respectively. Similar calculations were provided for a Si QD
with the same magnitudes of $\delta $, $B$ and $T$:
$\tau _{c}=3.5\times 10^{-9}$ s,
and $T_{2}=2.6\times 10^{6}$ s, $24$ s and $0.073$ s. Comparing this data
with the $T_{1}$ calculation of Ref.\cite{GlavinKim} shows that in spite of
the strong suppression of relaxation in the Si QD due to a small deviation $%
g_{0}-g$, the phonon-induced $g$-factor fluctuation via excited
states can control the phase relaxation (i.e., $T_{2}<T_{1}$) at $T\gtrsim 2$
K. The slight anisotropy of the $g$ factor\cite{WilsonFeher} results in some
$T_{2}$ dependence on the magnetic field direction. However, this effect is
expected to be small (less than 25\%) in contrast to the strong magnetic
anisotropy of $T_{1}$~\cite{GlavinKim}.

Qualitatively, another situation arises in the case of HFI modulation by the
phonon-assisted transitions. The distinctive feature of this mechanism is an
uncontrolled dispersion of the local nuclear field over the ensemble of QDs
due to the random distribution of nuclear spins. This dispersion
accounts for the fast (but partial) loss of initial electronic polarization
of the QD aggregate without spin coherence loss~\cite{KhaetLoss,SemKim}.
Thus, we take into account
the spin relaxation of a typical QD with a mean value of nuclear field
dispersion $\delta \Omega _{n}=a\sqrt{\frac{2}{3}I(I+1)\varkappa n_{I}/V_{QD}%
}$, where $I$ and $n_{I}$ are a nuclear spin and its concentration
in a QD of volume $V_{QD}$; the dimensionless parameter $\varkappa
=V_{QD}\int \left( \left\vert \Psi _{e}\left( \vec{r}\right)
\right\vert ^{2}-\left\vert \Psi _{g}\left( \vec{r}\right)
\right\vert ^{2}\right) ^{2}d^{3}\vec{r}$ is equal $9/16\pi $
within the approximations of Ref.\cite{GlavinKim}  and $a$ is a
constant in the HFI Hamiltonian $H_{HF}=a\vec{s}\vec{I}\left\vert
\Psi \left( \vec{R}\right) \right\vert ^{2}$ for the nuclear spin
$\vec{I}$ sited at a point $\vec{R}$ and an electron with its
envelope function $\Psi (\vec{r})$.
If we set $\Omega _{z}^{e}-\Omega _{z}^{g}=\delta \Omega _{n}$ in Eq.\ (\ref%
{eq8}), we readily find the following estimation%
\begin{equation}
T_{2,hf}^{-1}=\frac{\varkappa }{6}I(I+1)\frac{a^{2}n_{I}}{V_{QD}}\tau
_{\delta }F\left( \frac{\delta }{2T}\right) .  \label{eq14}
\end{equation}%
This equation shows independence of $T_{2,hf}$ on the magnetic field in the
case of HFI-induced transitions. An estimation of Eq.\ (\ref{eq14})
(with an appropriate HFI averaging over $^{69}Ga$, $^{71}Ga$ and
$^{75}As$~\cite{Lampel}) for a GaAs QD with a typical size of $50%
\mathop{\rm %
\text{\AA}}%
\times 500%
\mathop{\rm %
\text{\AA}}%
\times 500%
\mathop{\rm %
\text{\AA}}%
$ and $\delta =2$ meV gives $T_{2}=34$ s, $3.2\times 10^{-4}$ s
and $ 1.0\times 10^{-6}$ s for $T=1$ K, $2$ K, and $4$ K,
respectively. A similar estimation performed for a Si QD with the
same dimensions and temperatures results in $T_{2}=4.1\times
10^{7}$ s, $3.8\times 10^{2}$ s and $1.2$ s, respectively. One can
see that for the considered cases of both GaAs and Si QDs, the
contribution of HFI to phonon-assisted relaxation is small in
comparison to that of the $g$-factor modulation mechanism at $B=1$
T. However, the role of HFI modulation mechanism described in Eq.\
(\ref{eq14}) should prevail at weaker magnetic fields ($B<0.1$ T).

The common property of Eqs.\ (\ref{eq13}) and (\ref{eq14}) are their inverse
proportionality to the rate of phonon-induced transitions between electronic
states that qualitatively distinguishes this mechanism from those based
on the spin-flip transition probabilities~\cite{ErlNazFal}. This also
means that strong electron-phonon coupling may be favorable for slow
decoherence (in a manner analogous to the D'yakonov-Perel process in the
bulk materials). At the first glance, it would seem that $T_{2}^{-1}$ can be
enhanced limitlessly by increasing $\tau _{c}$. However, it will only be the
case when the inequality $\tau _{c}\ll T_{2}$ is not broken. An
increasing $\tau _{c}$ means that spin performs almost a full (or more)
revolution during the procession in the fluctuating field between the
sequential phonon-assisted electron transitions. Hence, in this a case we
expect the dependence of $T_{2}^{-1}\approx \tau _{c}^{-1}$
according to the physical nature of the mechanism under consideration.

Our formulas were obtained under the assumption of a two-level
electronic structure (with regard to the orbital degree of freedom).
In a similar
manner, one can consider a many-level case. For example, the
$N_{e}$ excited electron levels close together with an identical
$g$ factors or HFI can be taken into account by replacing the $%
\tanh x$ term with a function $\left( N_{e}e^{x}-e^{-x}\right) /\left(
N_{e}e^{x}+e^{-x}\right) $ in Eq.\ (\ref{eq12}).

When considering the HFI as a mechanism of spin relaxation, we restricted our
investigation to the SLR of single, but typical, QD with a mean configuration
of nuclear spins. It is important to bear in mind that the relaxation
measurements performed on a large ensemble of QDs correspond to averaging the
relaxation curves over random $T_{2,hf}$ instead of the single spin relaxation
with a mean $T_{2,hf}$ value. As a result, our mechanism would predict
non-exponential decay
of initial electron magnetization in a QD ensemble.

In conclusion, we considered an efficient mechanism of electron spin decoherence in
a single QD due to the fluctuating precession of the longitudinal (with
respect to direction of external magnetic field) effective magnetic field.
Compared to the SLR controlled by the spin-flip transitions, our
mechanism does not involve transitions between the Kramers doublets,
which leads to effective spin relaxation
characterized by  a smooth dependence on the applied magnetic field. On
the other hand, the mechanism under consideration reveal an exponential
dependence on the temperature when $T\ll \delta $. This provides an
opportunity to utilize the $T_{2}$ temperature
measurements as a tool for the study of QD's low energy structure. Although
detailed estimation of the spin relaxation rate depends on the
specific properties of the respective QDs,
the analysis on the typical cases illustrates an advantage of
Si QDs for the quantum computing applications due
to their very long relaxation time.

This work was supported in part by the Defense Advanced Research Projects
Agency and the Office of Naval Research.

\begin{figure}
\epsfig{figure=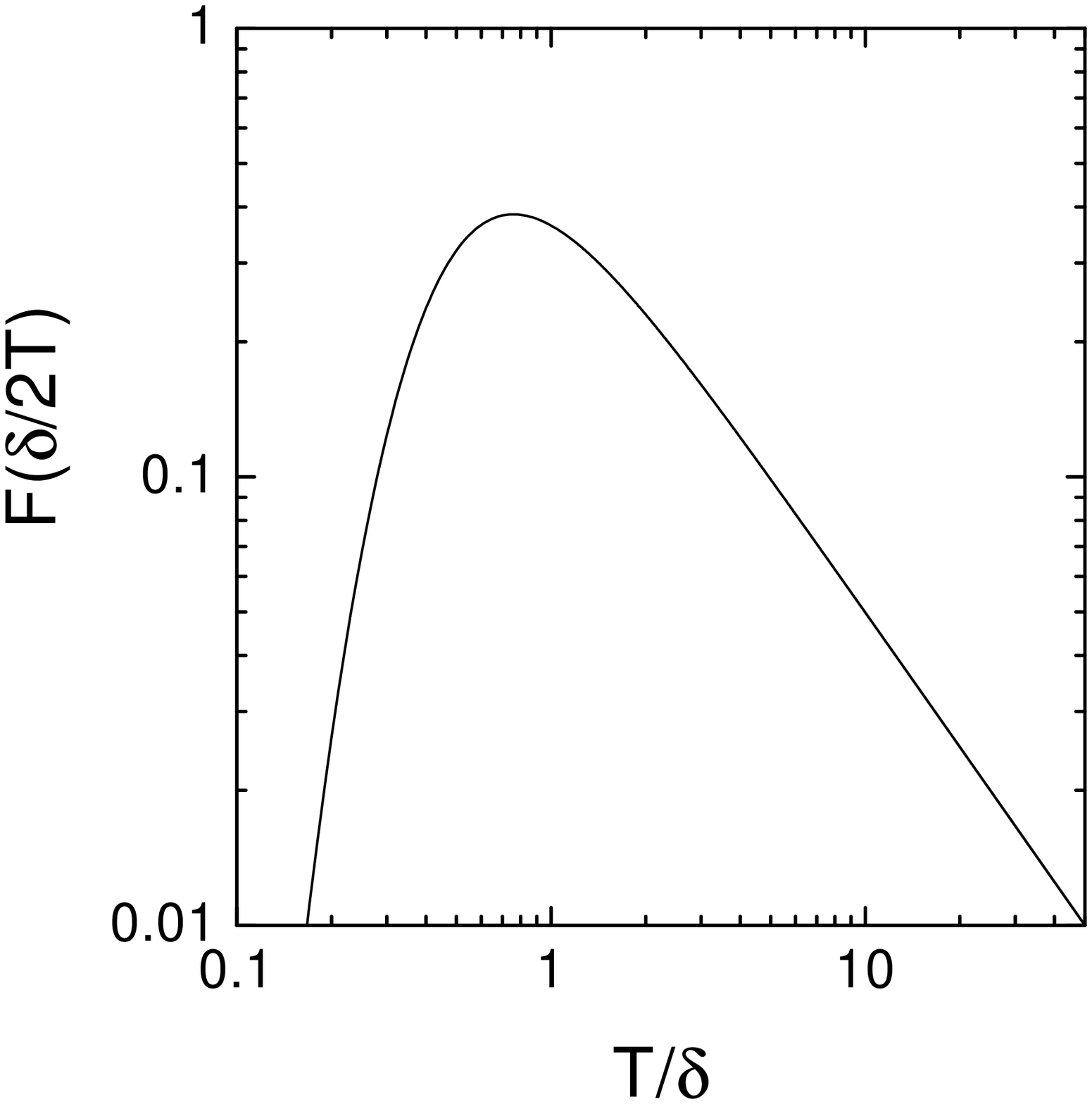,height=10cm,width=10cm,angle=0}
\caption{Decoehrence factor $F\left( \delta /2T\right) $ [Eq.\
(\ref{eq12})] as a function of temperature.}
\end{figure}

\end{document}